\documentclass[useAMS,usenatbib]{mn2e}
\usepackage{graphicx}
\usepackage{natbib}  
\newcommand\dm{D$_{\rm mix}$}

\newcommand{\msun}{\ensuremath{\, {M}_\odot}}
\newcommand{\Msun}{\ensuremath{\, {M}_\odot}}
\newcommand{\ocen}{$\omega$~Cen}

\title[Self--enrichment in Globular Clusters: role of the extreme anomalies]{A model for the Globular Cluster extreme anomalies}
\author[F. D'Antona and P. Ventura]{F. D'Antona$^{1}$\thanks{E-mail:
dantona@oa-roma.inaf.it (FD); ventura@oa-roma.inaf.it (PV)} and P. Ventura
$^{1}$\footnotemark[1]\thanks{This work has been supported through PRIN INAF 2005 
``Experimenting stellar nucleosynthesis in clean environments"
}\\
$^{1}$INAF, Osservatorio di Roma, MONTEPORZIO, I-00040, Italy
}
\begin{document}

\date{Accepted . Received ; in original form }

\pagerange{\pageref{firstpage}--\pageref{lastpage}} \pubyear{2006}

\maketitle

\label{firstpage}

\begin{abstract} In spite of the efforts made in the latest years, still there 
is no comprehensive explanation for the chemical anomalies of globular cluster 
stars. Among these, the most striking is oxygen depletion, which reaches values down to 
[O/Fe]$\sim -0.4$~ in most clusters, but in M13 it goes down to less than 
[O/Fe]$\sim -1$. In this work we suggest that the anomalies are due to the super position
of two different events: {\it 1) primordial self--enrichment}: this is asked to explain the
oxygen depletion down to a minimum value [O/Fe]$\sim -0.4$; {\it 2) extra mixing 
in a fraction of
the stars already born with anomalous composition}: these objects,
starting with already low [O/Fe], will reduce the oxygen abundance down to the most
extreme values. Contrary to other models that invoke extra mixing
to explain the chemical anomalies, we suggest that it is 
active only if there is a fraction
of the stars in which the primordial composition is not only oxygen depleted,
but also extremely helium rich (Y$\sim 0.4$), as found in a few GCs from their main
sequence multiplicity. We propose that the rotational evolution (and an associated extra mixing) of 
extremely helium rich stars may be affected by the fact that they develop a very small
or non existent molecular weight barrier during the evolution.
We show that extra mixing in these stars, having initial chemistry that has already
been CNO processed, affects mainly the oxygen abundance, and to
a much smaller extent if affects the sodium abundance. The model also predicts a
large fluorine depletion concomitant with the oxygen depletion, and a further
enhancement of the surface helium abundance, which reaches values close to Y=0.5
in the computed models.
We stress that, in this tentative explanation, those stars that are primordially O--depleted,  
but {\it are not} extremely helium rich do not suffer deep extra mixing.
\end{abstract}

\begin{keywords}
globular clusters; chemical abundances; self-enrichment
\end{keywords}

\section{Introduction}
The recent observations of abundance spreads among Globular Clusters stars, now 
observed also at the turnoff (TO) and among the subgiants 
\citep[e.g.,][]{gratton2001} show that these anomalies must be mainly attributed to 
some process of ``self--enrichment" occurring at the first stages of the life 
of the cluster. 
Two main sources have been proposed as basis for the self--enrichment:
\begin{enumerate}
\item {WMAGB: Winds of massive AGBs}. 
During the epoch starting some $\sim 5 \times 10^7$yr from the birth of the first stellar 
generation, when all the Supernovae have already exploded 
(carrying away from the clusters their high velocity ejecta) 
the massive Asymptotic Giant Branch (AGB) stars evolve, losing their envelopes
cycled through the hot CNO-cycle (Hot Bottom Burning --HBB) 
in low velocity winds, that may remain into the cluster and form the second generation
\citep{cottrell-dacosta, ventura2001}.
\item {WFRMS: winds of fast rotating massive stars}. 
The primordial gas cloud, from which the
first generation stars are born, is contaminated with the winds ejected from 
fastly rotating massive stars during the first few million years of the 
cluster life \citep{maedermeynet2006,decressin}. This model
implies a very fast second stage of star formation, preceding the supernova events
whose energetic input on the cluster gas may preclude any further star formation.
\end{enumerate}
In the case of WFRMS, the primordial gas from which the fist stellar generation is born
is certainly still present when the nuclearly processed stellar envelopes are ejected, and  
these latter may be diluted at different degrees in the primordial gas before new stars
are formed. In the case of WMAGB, 
remnant gas of the first generation may still be present and dilute the
ejecta either if the polluting stars are only the most massive AGBs (those referred to as super-AGBs),
or if, e.g., GCs are formed into dwarf galaxies 
at the time of the evolution of massive AGBs \citep{bekki2007}. In the latter case, the gas ejected
from the AGBs may be collected and pollute the gas present 
in the region where the cluster is being formed.
Otherwhise, it is scarcely credible that the winds 
are diluted with residual gas before forming stars of the second generation 
\citep{cottrell-dacosta}, as several decades of million years
have passes, and the supernovae have cleaned up the cluster. In fact, it has also been 
suggested that the spreads in chemical 
abundances are actually due to the birth of successive generation of stars {\it 
directly} from the ejecta of the massive AGBs of the first generation
\citep{dantona2002}. 

Neither model has been yet explored from a dynamic point of view.  
The role of the fast post-main sequence winds and of their energetic
input in the case of WFRMS is not clear. An escape from the fact that the supernova events will
be dramatic for the gas environment in the clusters is to consider that stars more massive than
$\sim$25\Msun\ may end their life collapsing into black holes. 
The advantage of this scenario is that the requirements on the initial
mass budget in massive stars is less severe than the constraints of the AGB scenario
\citep{pch06, decressin}.
In regard to the chemistry, both models are still not fully adequate, although there
is scarce doubt that the abundance anomalies we see are born in matter processed through the
hot CNO cycle, with little --if any-- influence of the third dredge up in AGBs \citep{ventura2002,
vda05a}.

An important prediction of both scenarios is that the second stellar generation must have helium
content enhanced, at some degree, with respect to the helium content of the first stellar
generation, that probably reflects the Big Bang abundance.
In recent years, the prediction of the spread that is naturally associated 
with the helium abundance in the AGB ejecta, allowed to
interpret the morphology of extended HBs in some globular clusters in terms of a spread 
in the initial helium content of the cluster stars \citep{dantona2002, 
dantonacaloi2004, lee2005}. Interestingly enough, after the \cite{dantonacaloi2004}
interpretation of the HB of NGC~2808 in terms of two populations differing in
initial helium content, a very peculiar helium distribution in the cluster
stars was inferred from the presence of a ``blue main sequence",
containing $\sim$15--20\% of stars, in the 
same cluster \citep{dantona2005}, similar to what had been found in 
$\omega$Cen \citep{bedin2004, norris2004, piotto2005}. New and better data by
\cite{piotto2007} have actually shown that the main sequence of NGC~2808 actually is
formed by three main components, precisely as the \cite{dantona2005} interpretation of the 
HB population required (see their Figure 9). In general, 
the main sequence morphology can allow to detect differences in helium content only when
the helium content is really very large \citep{dantona2002, salaris2006}. In fact,
both in NGC 2808 and  $\omega$Cen, a straightforward interpretation 
through stellar models assigns Y$\sim$0.4 to the bluest main sequence. This population however is 
only the ``tip of the iceberg", as the interpretation of the HB morphologies requires in
many clusters that 50\% or more of the stars have enhanced helium abundance 
\citep[e.g.][]{caloidantona2007, dantona2007}. 
The most straightforward interpretation of existing data is that most clusters 
have a population with enhanced helium (up to Y$\sim$0.3--0.33), but only a few have a small
population with {\it very high} helium, such as NGC~2808.
\begin{figure*}
\includegraphics[width=.48\textwidth]{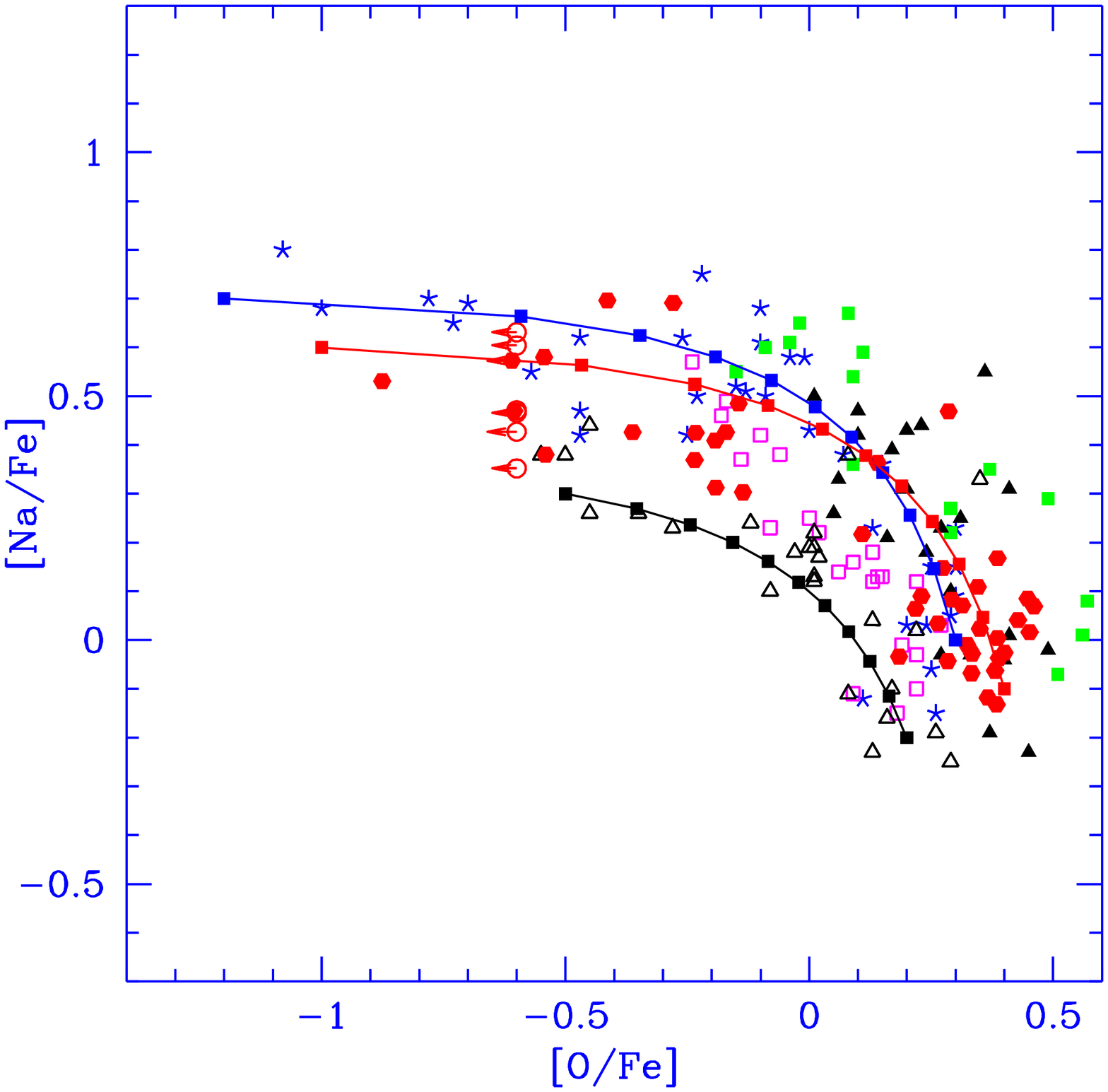}
\includegraphics[width=.48\textwidth]{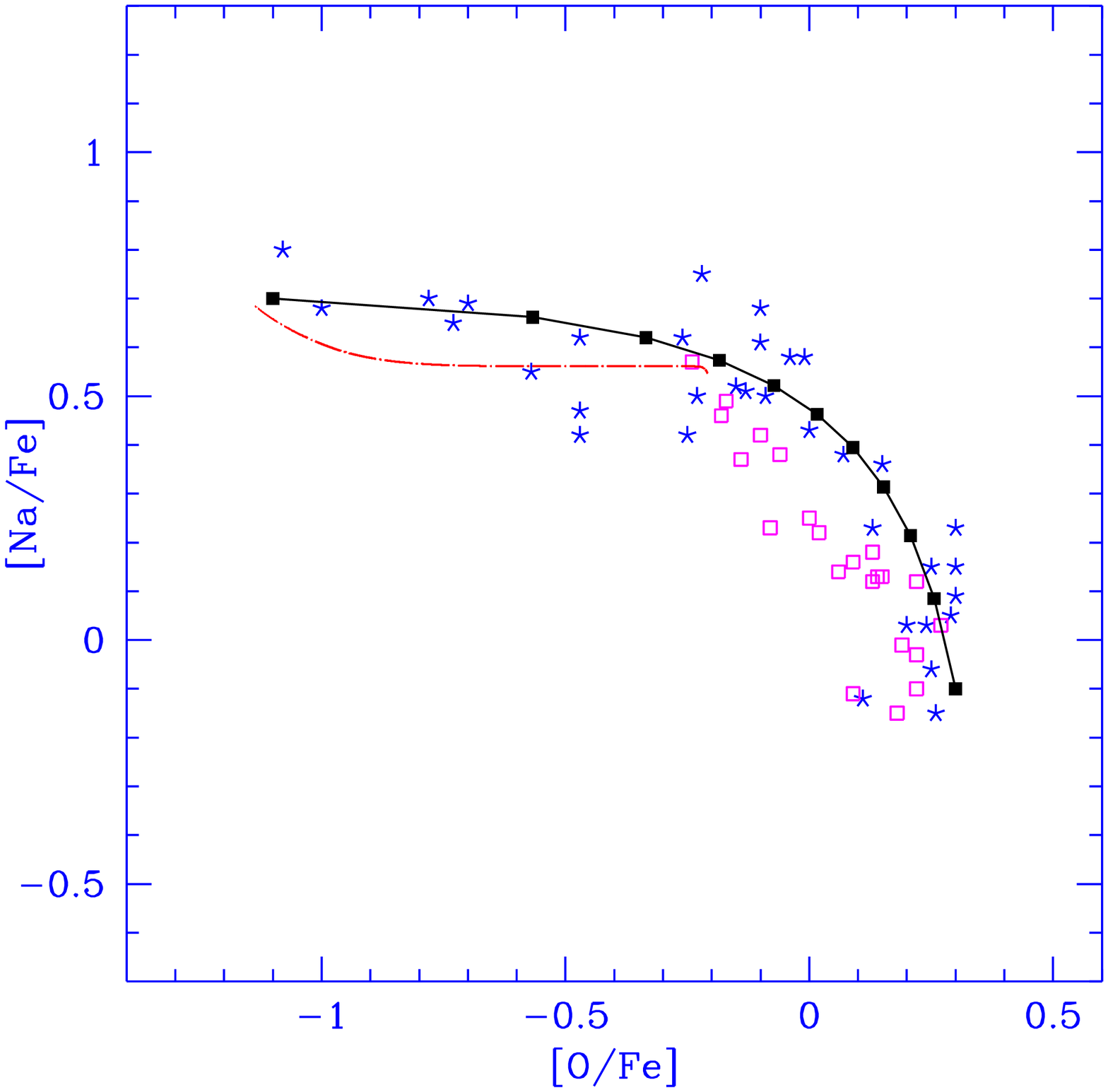}
\vspace{-50pt}
{ \caption{Left: observed data that define the anticorrelation sodium vs. oxygen in the stars 
of several GCs. stars: M13, open squares: M3 (both from Sneden et al. 2004); 
full squares: NGC 6752 from Grundahl et al. 2002; full triangles: M4 and open 
triangles: M5 (both from Ivans et al. 1999); full dots: NGC 2808 from 
Carretta et al. 2006. The three curves represent mixtures of different quantities of the two
compositions represented by the squares at each extreme. Going from left to right, the composition 
at the extreme left is diluted by 0, 10, 20 ... 90\% of mass having the composition at the
extreme right. Right: this figure shows only the M13 (stars) and M3 (open squares) data,
and the possible dilution curve for M13. The (red) dash--dotted line represent the result 
of extra mixing from the sequence 4 of Table 1.}}
 \label{f1}
\end{figure*}
We now try to understand whether we can recognize a similar behaviour
also in the sodium versus oxygen anticorrelation,
which represents the most clear evidence of chemical anomalies \citep[e.g.][]{sneden1991}.
This anticorrelation is present in most clusters examined so far. The maximum reduction 
found for oxygen in most clusters reaches a minimum value
[O/Fe]$\sim -0.4$\footnote{
as usual [X/Fe]=$\log ({\mathrm N}_X/{\mathrm N}_{Fe})-\log({\rm N}_X/{\rm N}_{Fe})_\odot$, where N 
indicates the abundances in number.}, while the ``normal" population in GCs and the
halo stars have [O/Fe]$\sim 0.3$
\citep[see, e.g.][~their Fig. 5]{carretta2006}. Sodium in the stars with 
reduced oxygen is in the range [Na/Fe]$\sim 0.3 - 0.7$, while [Na/Fe]$\sim$0 in the ``normal" stars.
The Na--O anticorrelation is also present among the TO and 
subgiant stars, as shown by \citet{gratton2001}, and therefore it can not  
inputed to `in situ' mixing. In addition to these typical anomalies, some
clusters have a small number of stars showing more extreme behaviour:
while the largest sodium abundance remains at $\sim$+0.7, oxygen
can be reduced down to  [O/Fe]$\sim -1.2$ in a few stars in the cluster M13
\citep[e.g.][]{kraft1993}. 
The cluster NGC~2808, which shows the ``blue main sequence" with Y$\sim$0.40 
comprising about 15\% of the cluster stars, also seems to have a similar fraction of 
extremely oxygen depleted stars, according to \cite{carretta2006}.
We are led to suggest, purely on observational grounds, 
that {\it the stars having the lowest oxygen values belong to the same group of 
those stars having the highest initial helium abundance}. 
Further, until today, the most extreme anomalies are only found in red giants.\\
In this work, we try to understand whether these information can be used
to derive some simple hint on the interpretation of the abundance anomalies, to be used 
as a constraint on the formation of the second stellar generation. 
In Section 2 we examine the possible
schemes for the explanation of the Na--O anticorrelation. Among the different
possibilities, we suggest that some among the ``typically extreme" ([O/Fe]$\sim$--0.4, 
[Na/Fe]$\sim$+0.5--0.7) stars evolve into the most extreme Oxygen poor giants 
due to an ``in situ" extra mixing. In other words, 
we will make the case that not all the abundance anomalies can be attributed to the chemistry 
processed in the progenitor stars (being either WMAGB or WFRMS pollutors), and that we need some
process of extra--mixing in the red giant stage to explain the most extreme anomalies.
In some clusters only, we suggest that there is extra mixing in red giants, and correlate it
to the existence of an extremely helium rich population. We
justify this suggestion on the basis of the different model structures during the RGB 
stage for differing stellar helium content (in Appendix). In Section 3 we summarize
the concept of ``canonical" and ``non canonical" extra--mixing in the literature. 
In Section 4 we model parametrically extra--mixing during the RGB 
evolution.  We show that the results of extra mixing depend on the
abundances assumed for the initial stellar chemistry.
Starting from already anomalous composition, deep extra mixing will mainly affect the oxygen 
abundance.
Table 1 summarizes the possible scenarios to understand the GC chemical anomalies.
We also show the effect of this parmetric extra mixing on the Fluorine abundance, and point out 
that the models, starting already with Y=0.4, show a surface helium increase up to Y$\sim$0.5,
which could be important to understand the very hot subdwarfs present in the HB of some clusters.

\section{Explaining the observational sodium vs. oxygen anticorrelation}
\subsection{The abundance patterns and the dilution model}
We show in Figure \ref{f1} the Sodium versus Oxygen abundances among the stars in 
several GCs. In contrast with the halo field stars, which occupy only the lower right
side of the figure (high oxygen, normal sodium), many of the GC stars show low oxygen
and higher sodium \citep[e.g.][]{sneden2004}. The case of the clusters M3 and M13 
is shown separately in the right side of the figure. 
The continuous lines with squares show a possible 
and straightforward interpretation of the abundance patterns: if we 
mix gas having the high Na--low O composition of the square at the extreme left with
``normal" matter having the low Na--high O composition represented by the square 
at the extreme right, we obtain all the intermediate compositions. 
The matter giving origin to the second stellar generation should then
have [O/Fe]$\sim$--1.2 and [Na/Fe]$\sim$0.7 in the case of M13, or less extreme values
for other clusters, e.g. [O/Fe]$\sim$-0.4 and [Na/Fe]$\sim$0.3 for the case of M5.

\begin{table*}                                                       
 \centering                                                          %
  \caption{Possible scenarios: composition of the polluting gas }       
  \begin{tabular}{@{}llccclc@{}}                                  
  \hline                                                                
Source & Cluster           & [O]   &   [Na]  & Y     & mechanism   & ? \\                                         
\hline
WFRMS      &   extreme GCs: & -1.3  &   0.7   & 0.40  & dilution with 1st gen. gas & not reliable: pollution by      \\ 
           &   (M13 ; NGC~2808)  &&&&&  fast winds is necessary\\
WFRMS      &   extreme GCs & -0.4  &   0.5   & 0.40  & dilution with 1st gen. gas, & \\
&&&&& extramixing for extreme c.c. & proposed here     \\ 		   
WFRMS      & less extreme GCs  & -0.4  &   0.5   & $\la$0.35  & dilution -- no extra mixing  & proposed here\\ 
\hline
WMAGB      & extreme GCs     & -0.4  &   0.5   & 0.40   & undiluted gas, or dilution with 1st gen. gas &proposed here \\
&&&&& extramixing for extreme c.c. &  \\   
WMAGB      & less extreme GCs  & -0.4 - 0.0  & 0.3 -  0.5   & $\la$0.35 & undiluted gas -- no extra mixing & \\   
\hline
\end{tabular}
\end{table*}

In addition to M13, extremely small oxygen abundances are possibly present also among 
the stars in NGC 2808, as shown by the upper limits of \cite{carretta2006} analysis.
For M13 and possibly for NGC~2808, we seem to need a site of nucleosynthesis 
(the envelopes of massive stars, or of AGB stars) in which the oxygen 
abundance is reduced by $\sim$1.5dex. 

\subsection{Models of FRMS}
The models by \cite{decressin} show that the matter lost in the winds has composition which
indeed keeps the signatures of CNO processing. For what concerns the Oxygen abundance, however,
the maximun reduction they find is from an initial abundance of 3.0$\times 10^{-4}$ to
$1.1\times 10^{-4}$\ for a 120\Msun\ evolution (their Table 5). At the same time,
the largest oxygen anomalies are found together with extreme helium overabundances (Y up to 0.5 
in the quoted 120\Msun\ evolution). These overabundance are of great importance, 
as we will see later on. Thus, starting from an 
$\alpha$--enhanced [O/Fe]$\sim$0.3 for the oxygen of the first stellar generation, 
the minimum abundance which could be found in second
generation stars would be [O/Fe]$\sim$-0.4. \cite{decressin} also consider the fact that the
Lithium abundances so far observed for turnoff stars in GCs are smaller than the primordial
Lithium abundance, but still quite large. They use the value 
A(Li)=1.93\footnote{the notation is A(Li)=log [N(Li)/N(H)]+12} by \cite{pasquini2005} 
to infer that the matter processed in the FRMS must be diluted with pristine matter (having A(Li)
=2.61, Coc et al. 2004) in the proportion 2/3 from WFRMS and 1/3 from the pristine matter, in
order to preserve this relatively low Lithium abundance. In this case, the minimum oxygen value
becomes [O/Fe]$\ge$--0.3. However the data on Lithium are scarce enough that we should postpone any judgment 
of this model based on Lithium to a more complete observational situation. 
There is an alternative possibility to reduce the oxygen content, which is worth mentioning: 
if we do not consider the average values for the abundances
in the winds of FRMS, but their time evolution, much smaller oxygen abundances 
(down to $\sim 7\times 10^{-6}$) can be achieved (Figure 6 in Decressin et al. 2006). 
If the winds may ``directly" form stars, some of them might be as extreme
as required by M13 observations. Notice however that, in this case, we need to use the chemistry
which the winds reach in the phase of He--burning. At this stage, the stars are no longer fast
rotating and the winds have high velocities, so that it is difficult to understand 
how they can be kept into the GC.

\subsection{Models for massive AGB stars}

For certain oxygen abundances as low as [O/Fe]$\sim$--1 are not achieved in 
the yields of intermediate mass AGBs computed until today \citep{ventura2006, bekki2007}.
Nevertheless,
recent modelling of intermediate mass AGB stars of low metallicity is not doing a bad job to 
provide a reasonable framework to explain a great part of che GC chemical anomalies.
Ventura \& D'Antona \citeyear{vda05a, vda05b} have considered carefully the role of
convection model and nuclear reaction rates in fulfilling the requirements of the
chemical patterns of abundances to be produced. 
In particular, they showed that the Na--O anticorrelation requires very efficient 
convection, to process the CNO through the ON cycle by HBB. A problem
arises as, at the same temperature 
necessary for destroying oxygen, also Na is efficiently burned \citep[e.g.][]{dh03}.
After some very parametric attempts to solve this problem by means of a very
small overshooting associated to the thermal pulse phase \citep{vda05c}, a
possible, very stable, solution is obtained by  lowering the reaction rate 
$^{23}$Na(p,$\alpha)^{20}$Ne within limits acceptable in the allowed limits for
this cross section, and raising by a factor $\sim2$\ the initial
neon abundance \citep{ventura2006}. Nevertheless, the most recent finding 
is that, for M$>5$\Msun, there is a sort of ``saturation"
in the combined effect of HBB and mass loss, and the oxygen reduction 
can not go beyond $\sim -0.7$dex\footnote{starting from [O/Fe]$\sim 0.3$,
this means that in these models [O/Fe]$\ge -0.4$, as in the case of
WFRMS.} (Ventura and D'Antona 2007, in preparation).
The sodium enhancement does not go beyond $\sim +0.5$dex, but this may be linked to 
the initial values assumed for the abundance of $^{22}$Ne.
These models are then able to explain the bulk of the Na--O anticorrelation, 
which is present in most GCs, but not the``extreme" stars of M13 and NGC~2808.

\subsection{A closer look at the data}

From 2.2 and 2.3 we conclude that both the possible sites for the nucleosynthesis
necessary to explain the chemical anomalies are not consistent with the very
low oxygen abundances of some giants of M13. Let us examine these in detail:
the data provided by \cite{sneden2004} for M13 contain 8 stars with  
[O/Fe]$<$-0.4 and [Na/Fe] from $\sim$0.4 to $\sim$0.8; among these, 
6 are brighter than Mv=-2.3 and are therefore essentially at the tip 
of the giant branch. Beside, 5 of them are surely variable, 
either semi- or irregular (I-48, II-90, 
L835, L940 from \citealt{kopacki2003}, L954 in \citealt{welty85}); 
one is a suspected variable (II-67 in \citealt{welty85}). 
For the two lower luminosity stars (III-59 and I-12) 
there are no indications of variability. The extreme abundances and (very low)
upper limits for oxygen concern a fraction of
the stars in NGC~2808 \citep{carretta2006}, but
also these data refer to evolved giants. The compilation by
\cite{carretta2006} of the available data for the Na--O anticorrelation shows 
clearly that the scarcely evolved stars in GCs  do not show abundances smaller than 
[O/Fe]$\sim$--0.4 and that the data below [O/Fe]=--0.4 all belong either to
M13 or to NGC~2808. In addition, although \ocen\ is certainly 
much more complicated than a typical GC, the sample of giants examined by
\cite{norrisdacosta1995} contains a few upper limits on oxygen
abundances below [O/Fe]$\sim -0.4$, for CN weak stars with
high sodium abundances. 
From this discussion,
it appears then possible that in {\it some} clusters, for
a fraction of the evolved stars, an {\it in situ} mechanism acts to produce the extreme
abundances. 

\subsection{Are the stars with extreme oxygen depletion also very helium rich?}

Let us assume that
the very helium rich sequences are present only in the clusters having the most extreme
anomalies. 
Then the minimum helium content of the processed matter
must be Y$\sim$0.4. 
As shown by \cite{decressin}, the winds of massive rotating stars are heavily enriched
in helium. In the dilution model, therefore, the most extreme
anomalies ---those associated with the pure stellar ejecta--- would also correspond to the
stars having the most extreme helium abundances. 
For the WMAGB model, the evolution of AGBs is
characterized on the one side by HBB, and on the other side by an efficient second
dredge up of helium, which raises the helium abundance in the envolopes to
values from Y$\sim$0.27 (for a 4\Msun star) to Y$\sim$0.3--0.32 for a 6\Msun.
There are until now no models computed for the super--AGBs (masses 7-8\Msun)
of population II, and we
can hypothize that these would reach the extreme helium required, Y$\sim$0.4.
Therefore, both models in principle suggest
that the maximum helium values are linked to the most extreme anomalies.
From an observational point of view, let us remember an intriguing result by
\cite{carretta2006} concerning NGC~2808. 
Dividing the cluster giants into three groups, with decreasing [O/Na], they 
notice that there is a very small increase in [Fe/H]
in the three groups, exactly as one would expect if the Na--O 
anomaly is associated with a larger and larger helium content of the studied
atmospheres: thus it is possible that the extremely low oxygen 
giants in NGC~2808 belong to the same group of stars which populate the blue main
sequence, and the extreme blue tails of the horizontal branch distribution.
Unfortunately, in M13, the only other cluster showing 
extremely low oxygen abundances, an high precision photometry of the MS is lacking,
to see whether a blue main sequence can be isolated. For \ocen, the other
cluster showing a blue main sequence which
can be interpreted in terms of very high helium
\citep{bedin2004,norris2004,piotto2005}, there might also
be giants with extreme values for the anticorrelation Na-O
\citep{norrisdacosta1995}, but further studies are needed.

We then have the following problem: the models may probably, in some cases, provide
ejecta with extremely high helium content, but section 2.2 and 2.3 suggest that this same matter
can not have the extreme oxygen content of M13 giants. These very low abundances, then, should
be due to some in situ mechanism (extra mixing) taking place only in these stars.
An important question we should ask is whether there is any physical reason to
link a very high initial helium abundance to extra--mixing during the RGB evolution.
We give a speculative answer in Appendix A. 

\begin{figure*}
\begin{center}
\includegraphics[width=.45\textwidth]{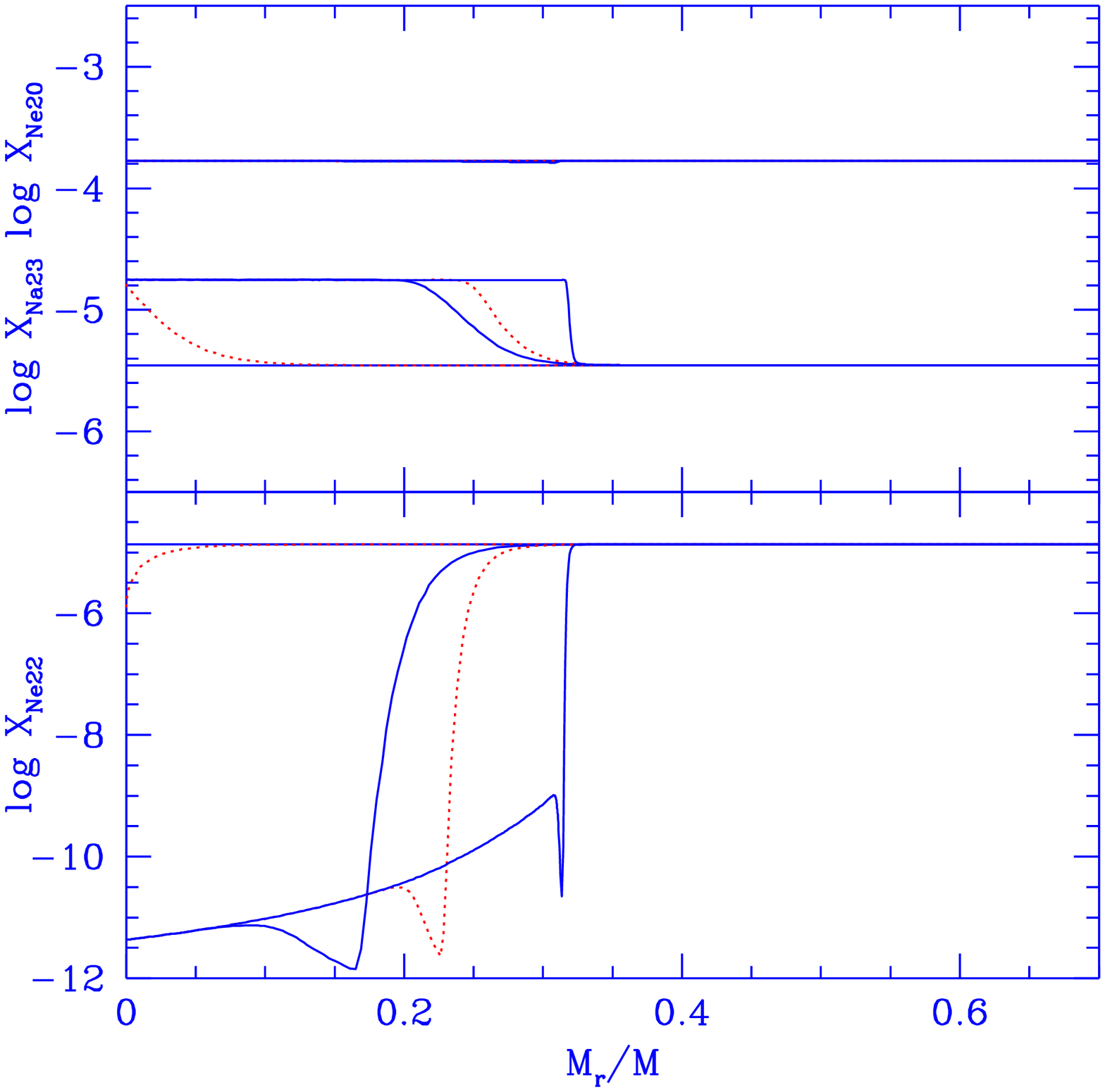}
\includegraphics[width=.45\textwidth]{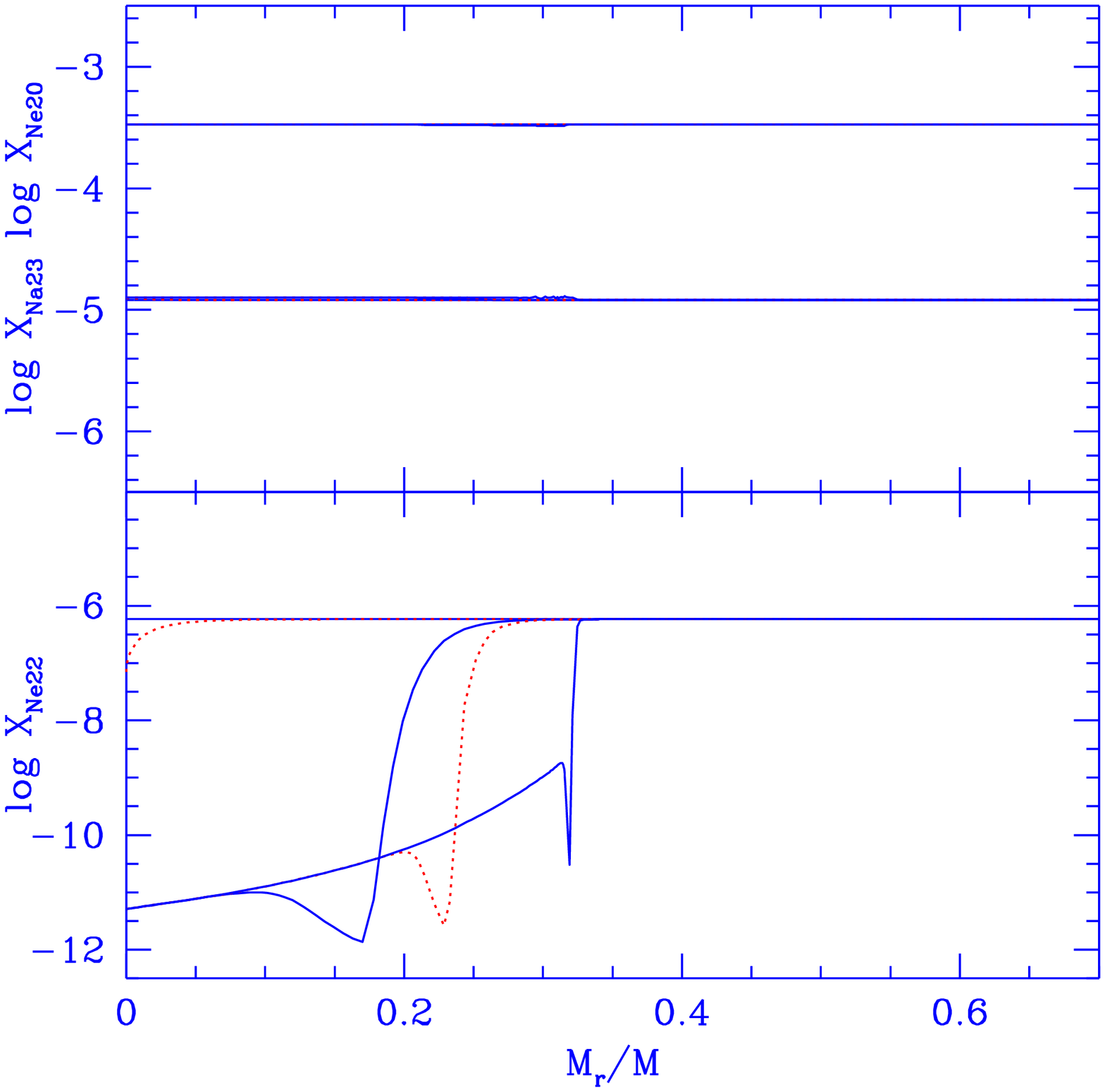}
\end{center}
\vspace{-50pt}
 \caption{The modification of $^{20}$Ne, $^{22}$Ne and $^{23}$Na profiles inside a red giant
 having solar scaled initial $^{22}$Ne (figures on the left) or depleted $^{22}$Ne (figures
 on the right). In the top panels, the upper curves (almost unchanged) refer to the $^{20}$Ne
 abundances, and the lower ones to the $^{22}$Ne. Alternate dotted and full lines are
 adopted to indicate successive evolutionary phases.
 The fractional mass is given as abscissa, the total mass is 0.64\Msun.
 In early phases of the RGB  evolution, the high $^{22}$Ne models burn 
 it to $^{23}$Na producing a profile of high sodium
 inside the giant. If the $^{22}$Ne is already low, the $^{23}$Na does not increase.}
\label{f2}
\end{figure*}

\section[]{Canonical and non canonical extra--mixing in red giants}

Standard mixing due to the convection description based on the Schwarzschild
instability criterion is not able to explain
the decline of the abundance of Carbon and
the strong reduction of the $^{12}$C/$^{13}$C isotopic ratio
in upper red giant branch (RGB) stars, both in the field and in star clusters
\citep{sneden1991}, and the decrease in surface Li 
abundance with increased luminosity \citep{pilach1993,grea00}.
Additional mixing seems to be efficient in the evolving
stars, as soon as the H-burning shell, advancing in mass, 
erases the chemical composition 
discontinuity left behind by the bottom of convective envelope at the end of 
the first dredge-up (\citealt{grea00,sh03}). At this moment, the evolution of 
red giants slows down for a while (see Fig. \ref{f3app}). 
This produces a prominent feature (the RGB 
bump) in the differential luminosity functions of globular clusters 
(\citealt{zea99,rea03}). The presence of the RGB bump can be taken as an 
experimental indication that the``standard" models of red giants actually make 
a good job in describing the evolution below the bump luminosity, 
and that no mechanisms of extra   
mixing in low-mass stars are acting, on the lower 
RGB, otherwhise the discontinuity left by convection would be smeared out and 
the bump would not be produced \citep{chaname}. The most credited hypothesis for 
preventing extra mixing below the bump is that the H-burning shell is shielded 
against mixing by the gradient in the mean molecular weight associated with the 
composition discontinuity (\citealt{sm79,chea98,dv03}) or, in any case, it 
operates very slowly (\citealt{chaname,pea06}). Very recently, \cite{eggleton2006}
proposed a different analysis: mixing arises between the supposedly stable and radiative 
zone between the hydrogen-burning shell and the base of the convective envelope, due 
to Rayleigh-Taylor instability within a zone just above the 
hydrogen-burning shell, where the burning of the $^3$He left behind by 
the retreating convective envelope {\it lowers} slightly the mean molecular weight, 
leading to a local minimum.  
\begin{table*}                                                       
\centering                                                          
  \caption{Evolution of chemical abundances in models with 
  \dm$=5 \times 10^9$cm$^2$s$^{-1}$}       
  \begin{tabular}{@{}lrccccccc@{}}                                  
  \hline                                                                
 \\
N& Nin & $\Delta$logT&[$^{12}$C]&[$^{14}$N] &[$^{16}$O] &[$^{19}$F]& [$^{22}$Ne] &[$^{23}$Na]  \\                
\\                                                                                                                                              
1  st &  1   & 0       &-0.862  &  0.904  &  -0.208 & -1.104 &    -1.367 & 0.539   \\                  
2  f   &600  &0.13   &-2.009  &  0.936  &  -0.209 & -1.106 &    -1.598 & 0.548    \\                  
3  f   &600  &0.05   &-1.353  &  1.128  &  -1.404 & -2.978 &    -2.336 & 0.761   \\                  
4  f   &600  &0.06   &-1.373  &  1.119  &  -1.140 & -2.740 &    -2.388 & 0.687   \\                  
5  ml  &600  &0.06   &-1.356  &  1.129  &  -1.467 & -3.047 &    -2.358 & 0.840   \\                  
6  f   &2000 &0.06   &-1.393  &  1.096  & -0.811  & -2.378 &    -2.416 & 0.672   \\                  
\hline
\end{tabular}
\end{table*}
 
\begin{table*}                                                       
\centering                                                          
 \caption{Evolution with different \dm and starting phase}        
   \begin{tabular}{@{}lrccccccccccc@{}}                                  
  \hline                                                                
 \\
N& N$_{in}$ & $\Delta$logT & \dm & [$^{12}$C]&[$^{14}$N] &[$^{16}$O] &[$^{19}$F]&[$^{22}$Ne] &[$^{23}$Na]  \\                
\\                                                                                                                                                           
1 st&  1  &    0  &    0    &-0.862  & 0.904 &-0.208  & 0.300  & -1.367 & 0.539          \\   
2 f &  600 &  0.06  & 5e9   &-1.373  & 1.119 &-1.140  & -2.74  & -2.388 & 0.687  \\                  
3 f &  600 &  0.06 &  5e8  & -1.534 & 1.077  &-0.656  &-2.279  & -3.205 & 0.581  \\   
4 f & 1050 &  0.06 &  5e8  & -1.555 & 1.055  &-0.537  &-2.121  & -3.160 & 0.578  \\   
5 f & 1050 &  0.06 & 2.5e8 & -1.576 & 1.033  &-0.446  &-1.872  & -2.737 & 0.568  \\   
6 f & 1500 &  0.06 &  5e8  & -1.287 & 1.015  &-0.405  &-1.590  & -1.954 & 0.568  \\   
7 f & 1500 &  0.06 & 2.5e8 & -1.085 & 0.980  &-0.327  &-1.349  & -1.646 & 0.557F  \\   
\hline
\end{tabular}
\end{table*}


For the majority of upper RGB stars, the surface abundance anomalies require that this
{\it in situ} mixing penetrates down to the outer part of 
the H-shell, where the CN branch of the CNO--cycle is operating. This is the
``canonical extra mixing'', in the definition of \citealt{dv03}.
These authors, parametrizing the depth and rate (diffusion coefficient) of extra-mixing below 
the formally convective region, find that
the values which describe the ``canonical" extra mixing are within the range from
$\Delta \log$T $\sim$ 0.19 and \dm $\sim 4 \times 10^8$ cm$^2 s^{-1}$ to
$\Delta \log$ T $\sim$ 0.22 and \dm $\sim 8 \times 10^8$ cm$^2 s^{-1}$. Here
$\Delta \log$T is the difference between the logarithms of temperature at the base of the
H-burning shell and at the maximum depth of extra mixing. These values then can explain
the evolutionary decline of [C/Fe] in the upper RGB stars, both in the field and in star
clusters. However,
canonical extra mixing does not affect O, Na, Mg and Al, the elements
which show anomalous abundances in the stars of GCs.
A ``non canonical'' extra mixing in upper RGB stars, penetrating the H-burning shell 
deeper than in the canonical case, could dredge up material deficient in O 
and enriched in Na (\citealt{dd90}),
These models constituted an alternative view to the
problem of GC chemical anomalies, until it was clear that these anomalies were present also 
in scarcely evolved stars. 
Once established that the self--enrichment due to star formation in matter either from
WFRMS and/or from WMAGB is a better scenario to explain globally the chemical
anomalies in GC stars, it remains to be understood whether mixing plays any additional role.
The possibility of enhanced, faster and deeper mixing,
has not been neglected as an {\it additional}
mechanism during the RGB evolution: e.g. \cite{dv03} have proposed that in some upper 
RGB stars canonical extra mixing may become deeper, driven 
by the differential rotation of the stellar radiative zones due, 
e.g., to spinning up of close binary members as a result of tidal 
synchronization (\citealt{dea06}). In addition, \cite{denis2006b} proposed
enhanced extra mixing to explain the Fluorine abundance variations in the red giants 
of the cluster M4.

In the following we will {\it assume} 
that indeed {\it a non canonical extra mixing is efficient only in most helium rich
fraction of cluster stars}, and we model its effect on the surface abundances. 
This combination of a primordial and deep mixing scenario has been invoked by
\cite{denis1998} to explain the magnesium vs. aluminum anticorrelation in GC red giants, and we
follow their line of reasoning, although we will devote to another study the comparison of 
the magnesium and aluminum abundances, and discuss here only the sodium and oxygen 
abundance behaviour.
Therefore, we will assume that the {\it initial} abundances of the evolving giants are
consistent with the chemistry obtained either in the WFRMS or in the WMAGBs. 
In order to have a precise choice for the computation of the models described below, we 
adopt the chemistry of the ejecta of a 5\msun\ star according to our AGB models
\citep{ventura2006}, but we assume the high helium abundance which would result from a higher mass
(Y=0.4).

\begin{figure}
\includegraphics[width=.48\textwidth]{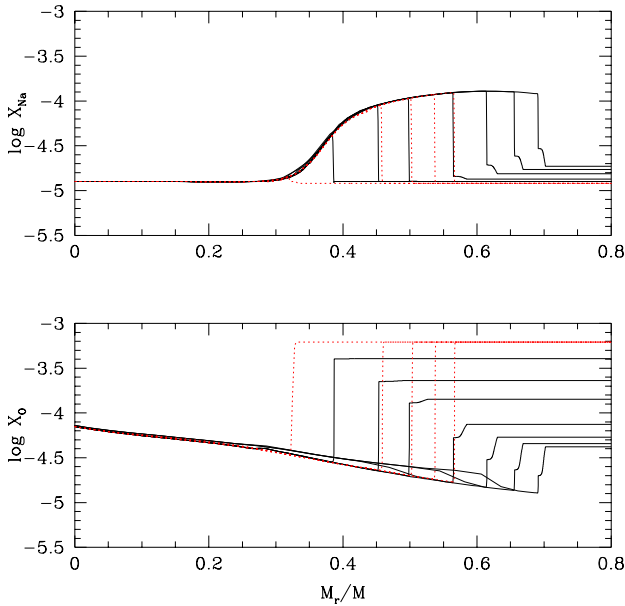}
\vspace{-50pt}
\caption{Sodium (top) and Oxygen (bottom) evolution in more evolved models of the 0.65\Msun.
The dotted lines refer to standard models, the continuous lines refer to models with deep 
extra mixing. 
Oxygen becomes more and more depleted at the shell boundary, by conversion to $^{14}$N, so that
fast extra mixing is at once effective in reducing its surface abundance. On the contrary,
only at high luminosity further burning of $^{22}$Ne finally establishes
a profile of $^{23}$Na into the star. Extra mixing can raise the surface abundance of
$^{23}$Na only in very late phases.}
\label{f3}
\end{figure}

\begin{figure}
\includegraphics[width=.48\textwidth]{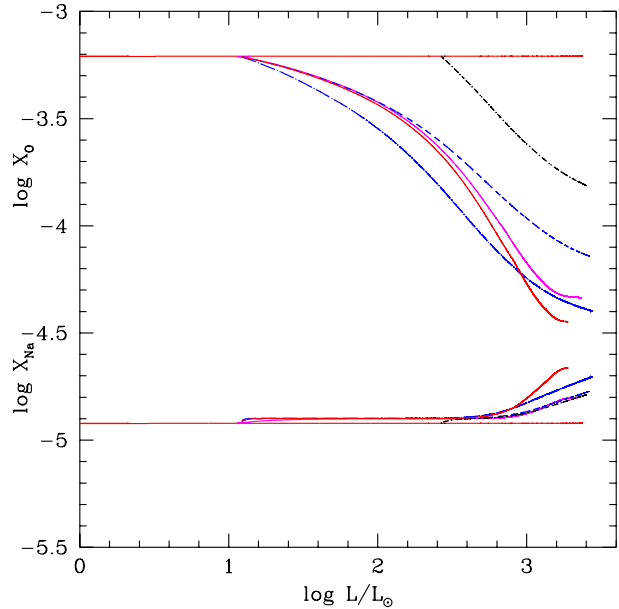}
\vspace{-50pt}
 \caption{Oxygen and sodium evolution as a function of the luminosity in the models of Table 2.}
 \label{f4}
\end{figure}

\subsection{Models for non canonical extramixing}

The models are computed by making use of the ATON code, described in its latest 
version in Ventura \& D'Antona \citeyear{vda05a, vda05b}. 
We compute standard models, in which mixing is purely due to convection, and
no kind of overshooting is allowed below the external convective region.
A description of these models is given in Appendix A. 
Here we describe the models with extra mixing, which refer to M=0.64\Msun\ with
initial helium mass fraction Y=0.4, and metallicity in mass fraction Z=0.002.
The helium abundance is selected
following our reasoning which links the extra mixing to the initially
very high helium abundance both as a consequence of observations (Section 2) and
of the interpretation of the RGB evolution as a function of Y (Appendix A).
As we will justify in the Appendix, we do not link the occurrence of extra mixing
to a particular stage of evolution along the RGB, but we consider different
initial stages, from luminosities at the beginning of the RGB evolution, up to
luminosities larger than those of the (very mild) RGB bump for the chosen Y.
The mass 0.64\Msun\ is chosen as representative of a mass presently evolving in a typical 
GC (age of $\sim 12$Gyr) for Y=0.40 and Z=0.002.

Some model results are listed in Table 2 and 3. 
At this stage, we do not aim to understand what is the specific mechanism
for extra mixing, but only which are the consequences for the surface abundances of
the giants presently evolving in GCs.
We adopt the description by Denissenkov \& VandenBerg (2003), and assume those
depths and rates of extra-mixing, below the formally convective region, necessary to
obtain sensible changes of the surface abundance of oxygen. Values of 
$\Delta \log$T $\sim$ 0.05 -- 0.06 and \dm $\sim 0.1 - 5 \times 10^9$ cm$^2 s^{-1}$
are adopted.
We also followed an evolution by considering the mass loss by stellar wind. 
Table 2 lists in the first column an identification sequential number and the
phase of evolution is represented. Sequence 1 (`st') is the sequence computed without extra mixing,
while `f' means that we provide the abundances at the end of the computed sequence, 
corresponding to the onset of the helium flash. In the `ml' sequence 5 of Table 2, 
mass loss is included in
the computation, according to the description by Reimers': 
\begin{equation}
\dot{M}=4.83 \times 10^{-9} \eta_R LR/M
\end{equation}
where L, R and M are respectively the stellar luminosity, radius and mass in solar units and
$\eta_R$\ is the free parameter. In this case we take $\eta_R$=0.29.

The third column lists the initial model at which extra mixing starts, beginning from 
the sequence with no extra mixing. The luminosity, T$_{\rm eff}$~ and helium core mass 
corresponding to the model numbers are listed in Table 4.
The following columns provide the logarithm of abundance of the listed elements, with respect to 
the standard solar--scaled abundances. The values at model 1 of the standard sequence 1 do not 
change along the evolution, so these element abundances also represent the starting abundances of the
other sequences.

\begin{table}                                                       
 \centering                                                          %
  \caption{HR diagram location and helium core of models along the standard 0.64\Msun sequence }       
  \begin{tabular}{cccc}                                  
  \hline                                                                
N   & logL/L$_\odot$ & log T$_{\rm eff}$ & M(H)/M$_\odot$  \\                                         
500 & 0.567  & 3.752 &0.099 \\
600 & 0.909  & 3.727 & 0.165    \\ 
900 & 1.604  & 3.692 & 0.234  \\  
1050 & 1.797 & 3.681 & 0.253 \\   
1500 & 2.147 & 3.660 & 0.291 \\   
2000 & 2.427 & 3.641 & 0.319    \\  
\hline
\end{tabular}
\end{table}
 
\section{Chemical stratification inside the RGs and the effect of the initial chemistry}

The effect of deep mixing depends on the profiles established within the star for the elements into
consideration. These profiles depend on the one side on the temperatures reached within the
inner layers during the evolution, and on the other side on the {\it initial} abundances of these 
same elements and of those which produce them.
For the elements which are our main concern now, in particular, oxygen and sodium, in fact, we
wish to study the effect of deep extra mixing starting from the initial mixture that comes out
from the evolution of a 5\msun\ of metal abundance in mass fraction Z=2$\times 10^{-3}$, according 
to the models described in \cite{ventura2006}, These abundances are listed in the first line
of Tables 2 and 3. In particular, Oxygen is already depleted by -0.508 dex with respect to its initial abundance
([O/Fe]=0.3)\footnote{The choice of an initial abundance [O/Fe]$\sim -0.2$\ may be confusing, as
we have stated that abundances as low as [O/Fe]$\sim -0.4$\ may be achieved in massive AGB models.
We preferred to use ``directly" the abundances coming out from a precise stellar evolution, namely
that of the 5\Msun, as we have not yet completed the study of the evolution of more massive models
which have the maximum oxygen depletion (Ventura and D'Antona 2007, in preparation).}, Sodium 
is already enhanced by +0.54dex, and $^{22}$Ne is depleted by -1.37dex.

These initial abundances have an obvious but subtle effect on the result of extra mixing. In particular,
let us first consider the  $^{22}$Ne and  $^{23}$Na abundances. Let us consider 
models starting from solar scaled abundances: The  $^{22}$Ne abundance is initially almost 4 times
larger than the  $^{23}$Na. There is ample space for  $^{22}$Ne to be burned inside the giant and
establish a large gradient of  $^{23}$Na across the burning shell, even at relatively low luminosities.
We show in the left side of Figure \ref{f2} the formation of this sodium gradient at several
stages during the standard giant evolution of a 0.64\Msun\ having solar scaled initial abundances of 
these two elements. The advancing shell burning shifts to larger fractionary masses the formation
of high sodium and the burning of  $^{22}$Ne by proton captures. But in the AGB 
envelopes which we assume to be progenitors of the giants which suffer extra mixing, 
already a great fraction of $^{22}$Ne had been converted to $^{23}$Na by HBB. 
Therefore, during the first phases of RGB evolution, no further $^{23}$Na is produced,
and no sodium profile is established in the star, contrary to
the case of abundances in solar ratios (right side of Figure \ref{f2}). 
It is only later on during the RGB evolution, 
when the shell temperature is higher, that $^{23}$Na is again produced at the expenses of the
burning of a very small fraction of the abundant
$^{20}$Ne. This is shown in the top of Figure \ref{f3}: sodium enhancement 
at the border of the burning shell can be obtained only when the core mass becomes larger than
$\sim$0.3\Msun, that is at high luminosity. In this figure the dotted lines represent the sodium 
(top) and oxygen (bottom) profiles in the standard model evolution. The abscissa, as usual, is the
fractionary mass. The location of the abundance gradients corresponds to the core mass (and therefore
luminosity --see Table 4) achieved at each model. We see that, while sodium is
produced and oxygen depleted inside the star, nothing happens to the surface abundances
if there is no extra mixing. The continuous lines show the evolution of the 
chemical profiles into some models of sequence 4 of Table 2. We see that oxygen
begins to be depleted at much smaller core masses (luminosities) than those at which some effect on the
increase of sodium begin to occur. 
Notice that we have started from an {\it already depleted} oxygen abundance: if we had started from 
solar scaled --or, better for standard population II, enhanced-- oxygen, the dilution effect would have
been much larger.
In fact, in the {\it classic} interpretation of abundance anomalies by deep mixing, starting from
standard initial abundances is possible to interpret the whole sodium vs. oxygen anticorrelation
in term of deep mixing \citep[see, e.g.,][ their figure 9]{weiss2000}.  

The fact that the effect of extramixing is more important for
the oxygen depletion than for the sodium increase is very clear by looking at
Figure \ref{f4}, which shows the abundances of oxygen
and sodium vs. the stellar luminosity, for the extra mixing models detailed in Table 2. One
of these results (model 4) is shown in the right side of Figure \ref{f1}. We see then that in
fact the very low oxygen abundance of M13 giants can be consistent with an in situ extra mixing
starting at non extreme abundance anomalies ([O/Fe]$\ge$--0.4, [Na/Fe]$\sim$+0.5) 
provided by the massive AGB models.
\begin{figure*}
\begin{center}
\includegraphics[width=.45\textwidth]{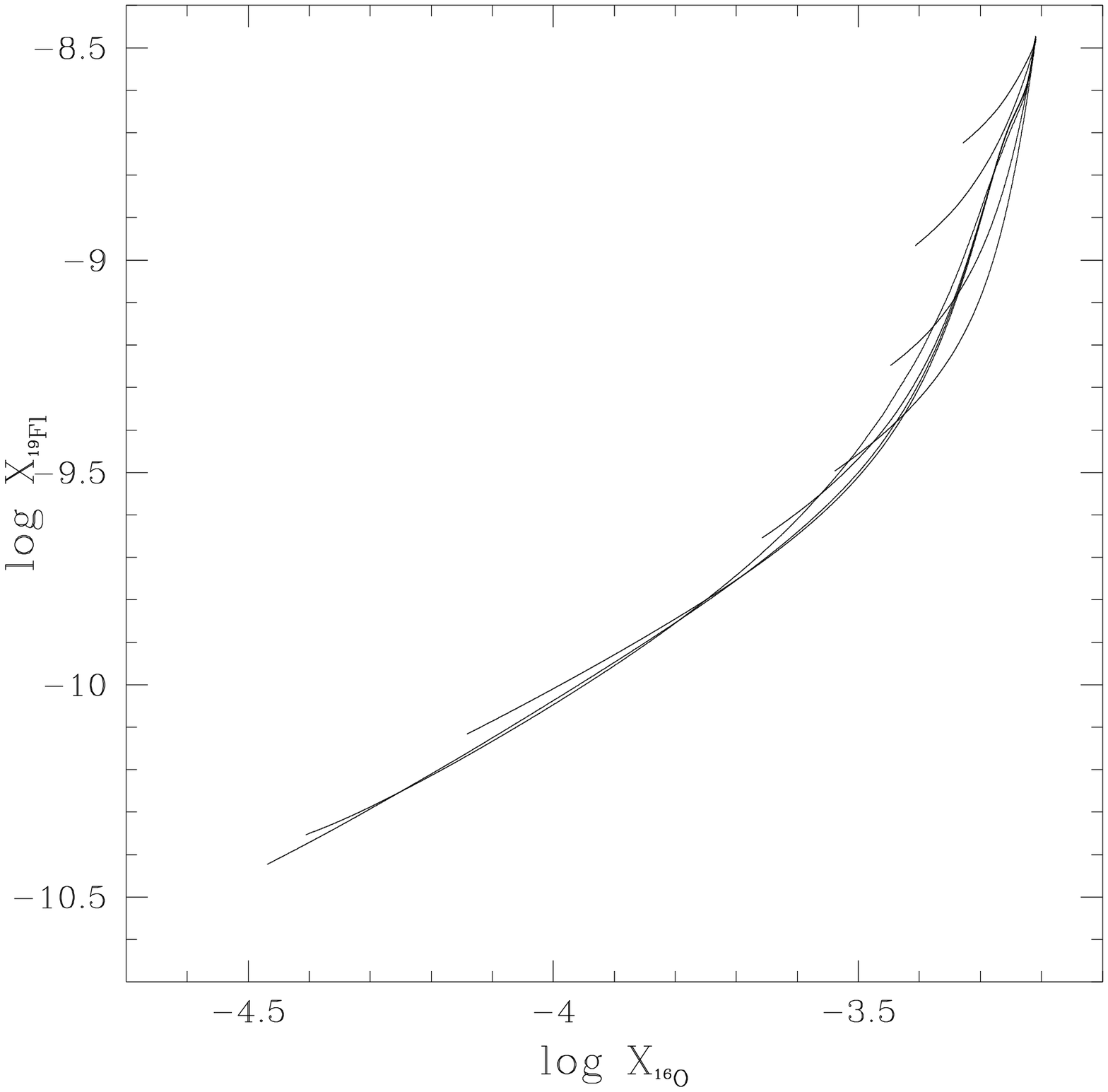}
\includegraphics[width=.45\textwidth]{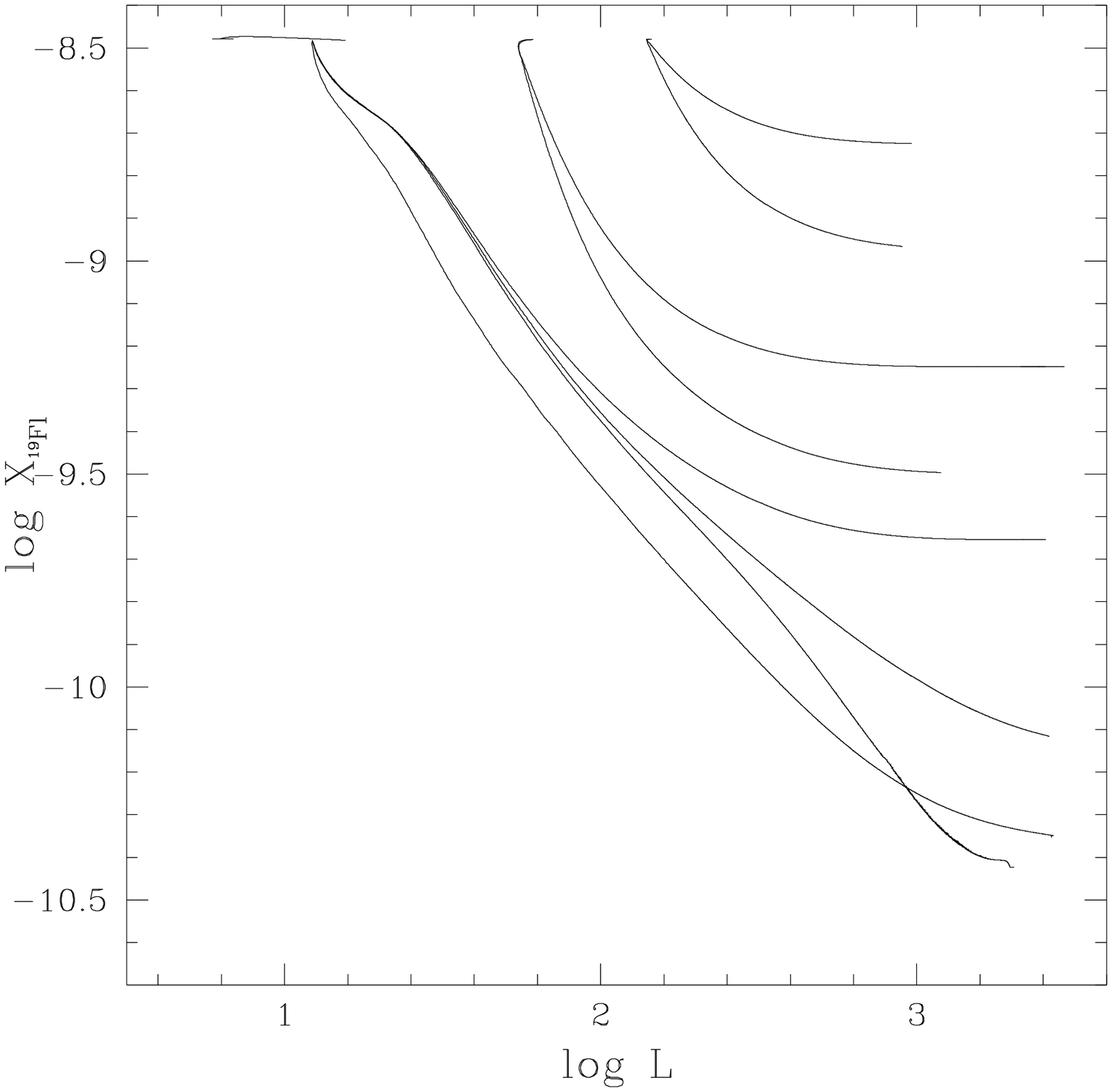}
\end{center}
\vspace{-10pt}
 \caption{The modification of $^{19}$F abundance for various evolutions (Table 3) are
 shown as a function of the oxygen abundance (left) and or the stellar luminosity (right figure).
 }
\label{f5}
\end{figure*}

\subsection{The Fluorine abundance}
Fluorine is already depleted by a factor $\sim$10 by HBB in the 5\Msun\ evolution from which we take the
initial abundances. This depletion is consistent with the data by \cite{sea05} for the giants of
M4. This cluster should not be as extreme as M13 or NGC~2808 which we are considering here, so
the F vs. Na anticorrelation should be a result of the primordial abundance variations, which
are predicted by our models. Nevertheless, \cite{denis2006b} point out that \cite{sea05} data also
show an average decrease of the fluorine and carbon abundance for increasing luminosities of the M4 giants,
and attribute the anticorrelation to extra mixing, explored with models similar to those we are
presenting here.
In our models with extra mixing, fluorine is in fact 
heavily depleted, especially in the models having the  
largest diffusion coefficient. Figure \ref{f5} shows the fluorine vs. oxygen anticorrelation and allows
a prediction: in the most extreme giants of M13 and NGC~2808, having the lowest oxygen abundance which
we attribute to extra mixing, the fluorine abundance must be more or less reduced depending on  
the fastness of extra mixing.
Table 3 shows in fact that the diffusion coefficient is an important ingredient 
to achieve strong depletions (see the comparison between models 4 and 5 or the comparison
between models 6 and 7).

\begin{table}                                                       
 \centering                                                          %
  \caption{Final helium mass fraction in models of Table 1 }       
  \begin{tabular}{@{}llccc@{}}                                  
  \hline                                                                
N     & Nin & $\delta$logT  & Y\\                                         
1 st & 1   & 0        & 0.404   \\ 
2 f   &600  &0.13      & 0.424  \\  
3 f   &600  &0.05      & 0.464  \\   
4 f   &600  &0.06      & 0.450  \\   
5 ml    &600  &0.06      & 0.482  \\   
6 f   &2000 &0.06      & 0.422   \\  
\hline
\end{tabular}
\end{table}

\section{The final helium abundance in the envelope}
Extra mixing affects the envelope helium abundance \citep{weiss2000}. In our case, the
envelope abundance is already a high Y=0.4, and it increases up to Y=0.48 in the most ``realistic"
sequence, computed with mass loss. We do not explore further
this problem here, but mention that some of our models also showed higher envelope helium
abundances. This fact has important consequences for the location of
these models on the following Horizontal Branch phases. The very hot HB stars at 
T$_{\rm eff}>31000$K can not be explained by standard HB models, even with very low 
hydrogen envelope on top. Sweigart (1997) suggested that a late helium
flash along the white dwarf cooling sequence can penetrate the
small entropy barrier of the weakly burning hydrogen-burning
shell (Iben 1976), so that helium and carbon would come to the surface, and hotter 
T$_{\rm eff}$'s can be achieved. We see in the models with extramixing 
that, maybe, the very high envelope Y is a result of deep mixing in giants which {\it already}
from the beginning have a high helium abundance, and suggest that the existence of
a hot blue tail in the HB may be linked to the extremely high helium abundances in the envelope,
a combined result of the {\it initially} already high helium and of deep extra mixing. 
The evolution of these stars will be described in a subsequent paper (Caloi et al. 2007, 
in preparation).

\section{Conclusions}
The chemical anomalies in GCs include both oxygen depletion and helium enrichment.
While there is no totally accepted model for the site of nucleosynthesis which affected
the matter from which the anomalous stars were formed, it is accepted that helium enrichment
will go together with oxygen depletion. Helium enhancements up to Y$\sim$0.4, as seen in a few
main sequences of GCs, 
may result from models, while the same models are easily consistent with 
the depletion of oxygen down to [O/Fe]$\sim$--0.4 shown by most GCs. The models
on the contrary have serious difficulties with the 
much larger depletions which are seen in some giants in M13, or may be inferred by the
spectral analysis of the giants in NGC~2808. 

In this work we remark that the evolution 
along the red giant branch (RGB) for stars having Y$\sim$0.4 
is noticeably different from the evolution of 
normal red giants with Y$\sim 0.24$: in particular, both the milder external 
convection penetration, and the higher initial helium abundance in the envelope 
conspire to produce a smaller discontinuity in molecular weight along the 
evolution (in particular, the standard "red giant bump", occurring when the 
hydrogen burning shell crosses the molecular weight discontinuity, is less 
prominent and occurs at a larger luminosity). 

We suggest that only the stars having very high helium abundance (if they are indeed
present in the GC) are subject to deep non standard extra--mixing below the 
formal convection zone, a mechanism which is not efficient in the presence of 
the molecular weight barrier. We propose that extra mixing may be efficient along 
the whole RGB evolution, if the helium content is large enough to minimize the 
molecular weight discontinuities, and attribute to it the occurrence of the extreme 
abundance anomalies which are present in a few GCs. 

We model extra mixing then starting from {\it already anomalous} chemistry, and show that
it affects mainly the oxygen --and not the sodium-- abundance. We also predict a very 
strong depletion of Fluorine, and helium envelope abundances as high as Y$\sim$0.48 in
present models. 

These results may serve to clarify the role of primordial self--enrichment in GCs, showing that
self--enrichment should not necessarily explain also the most drastic chemical compositions found in
some clusters.

\section*{Acknowledgments} 
We thank Vittoria Caloi, Raffaele Gratton and Thibaut Decressin for enlightening conversations.

\appendix

\section{Modelling the evolution of helium rich models along the RGB}

\begin{figure*}
\begin{center} 
\includegraphics[width=58mm]{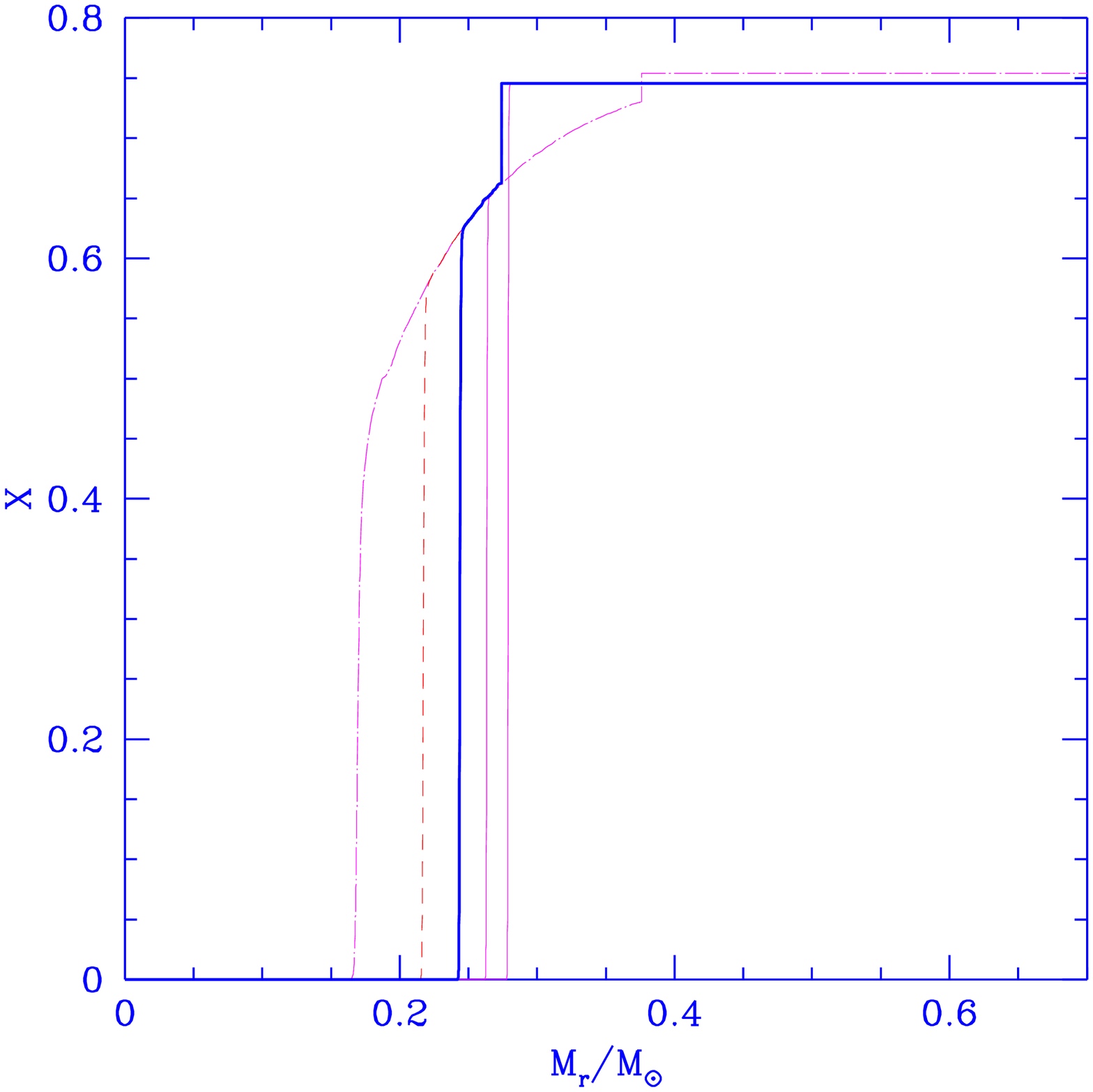}
\includegraphics[width=58mm]{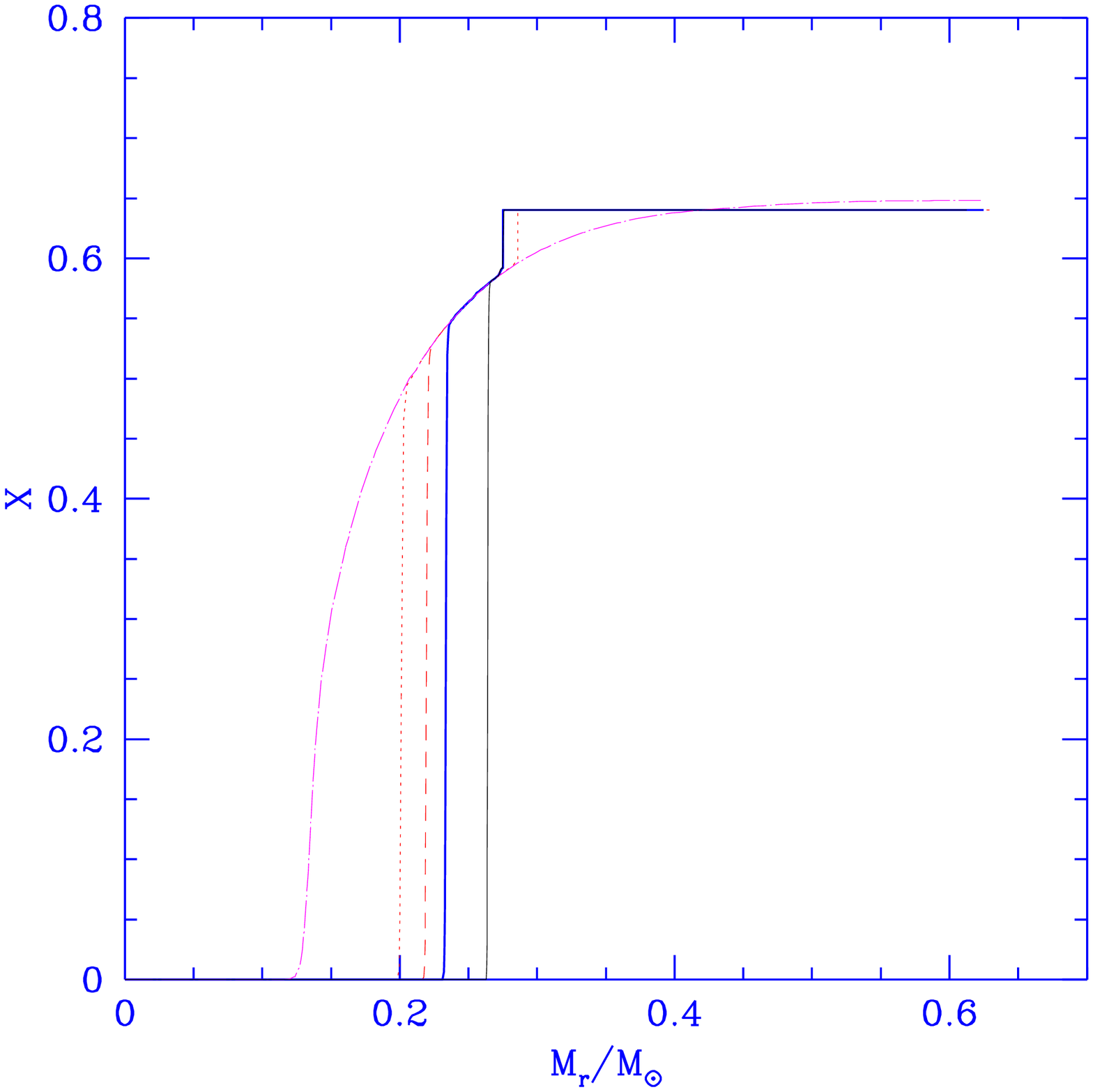}
\includegraphics[width=58mm]{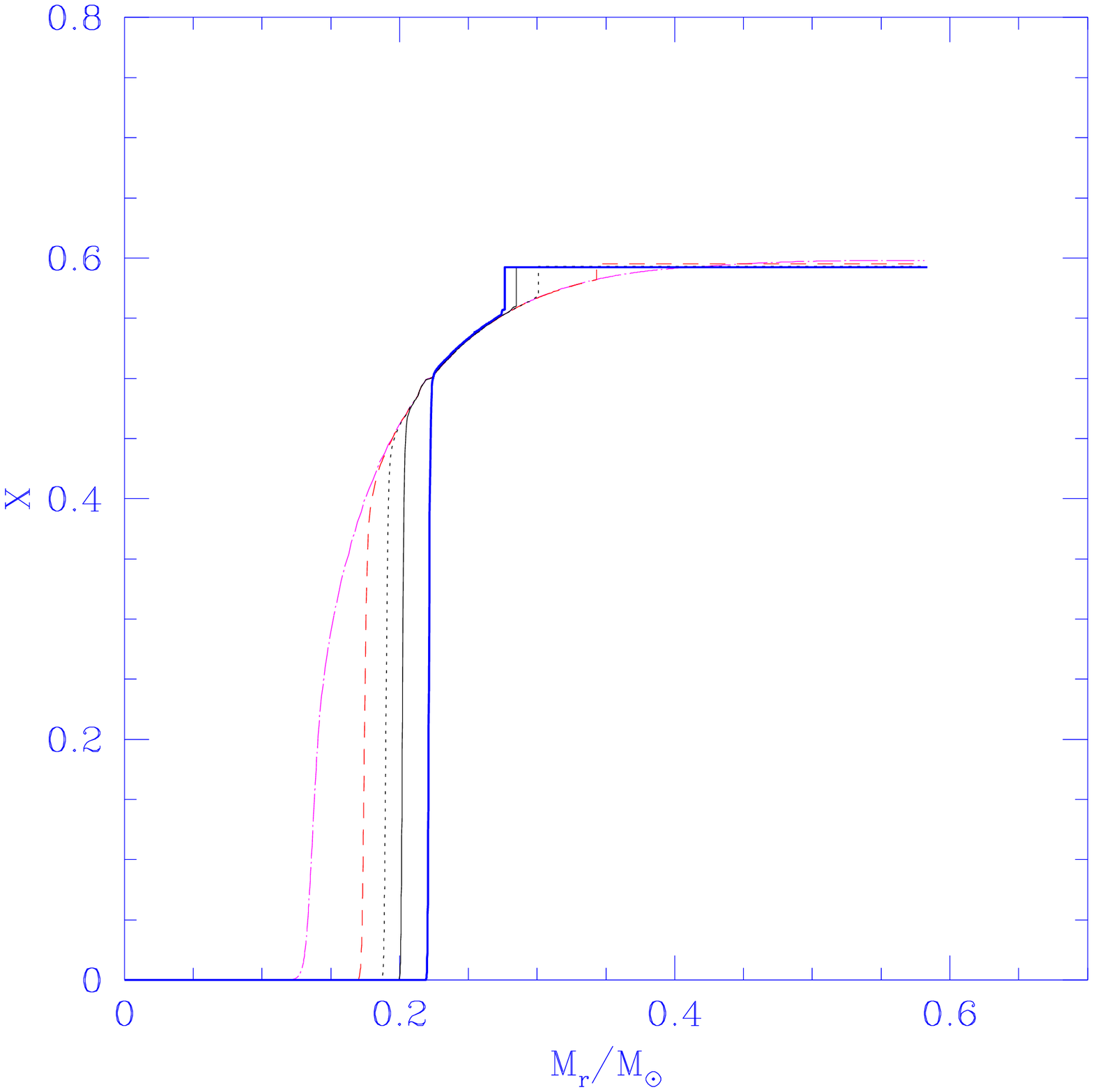} 
\end{center}
\vspace{-40pt}  
\caption{From left to right we show the evolution of the hydrogen profile along the
three evolutions of Figure A1. The chemical discontinuity is very evident in the
Y=0.24 case (left), much smaller in the Y=0.35 case (center) and very small in the 
Y=0.40 case (right).
}
\label{f1app}
\end{figure*}

\begin{figure}
\includegraphics[width=.48\textwidth]{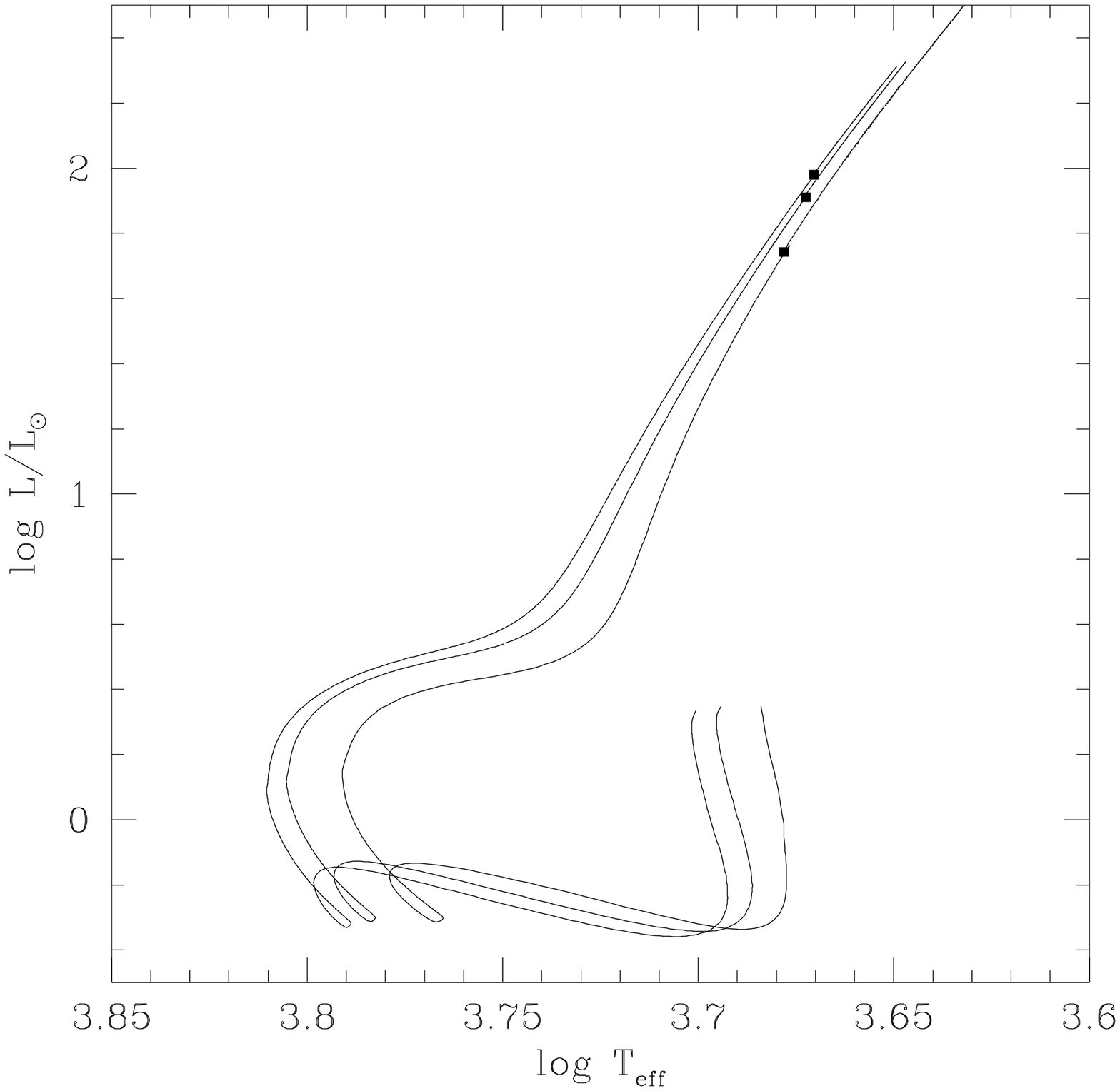}
 \caption{HR diagram. From right to left are shown the 
0.8\Msun~ (Y=0.24), 0.7\Msun~ (Y=0.35) and 0.64\Msun~ (Y=0.40)
for Z=$2 \times$10$^{-3}$. The three dots on the tracks show the location of the red giant bump.}
\label{f2app}
\end{figure}

We describe here the evolution of three
masses having different helium content: Y=0.24 for M=0.8\Msun, Y=0.35 for M=0.7\Msun
and Y=0.4 for Y=0.4\Msun, and a metallicity in mass fraction Z=0.002.
The masses are chosen as representative of masses presently evolving in GC (age of
$\sim 10-12$Gyr).
We compute standard models, in which mixing is purely due to convection, and
no kind of overshooting is allowed below the external convective region.
We show in Fig.\ref{f1app} the hydrogen abundance along the RGB evolution of 
The corresponding HR diagram is shown in Fig.\ref{f2app}.

\begin{figure}
\includegraphics[width=.48\textwidth]{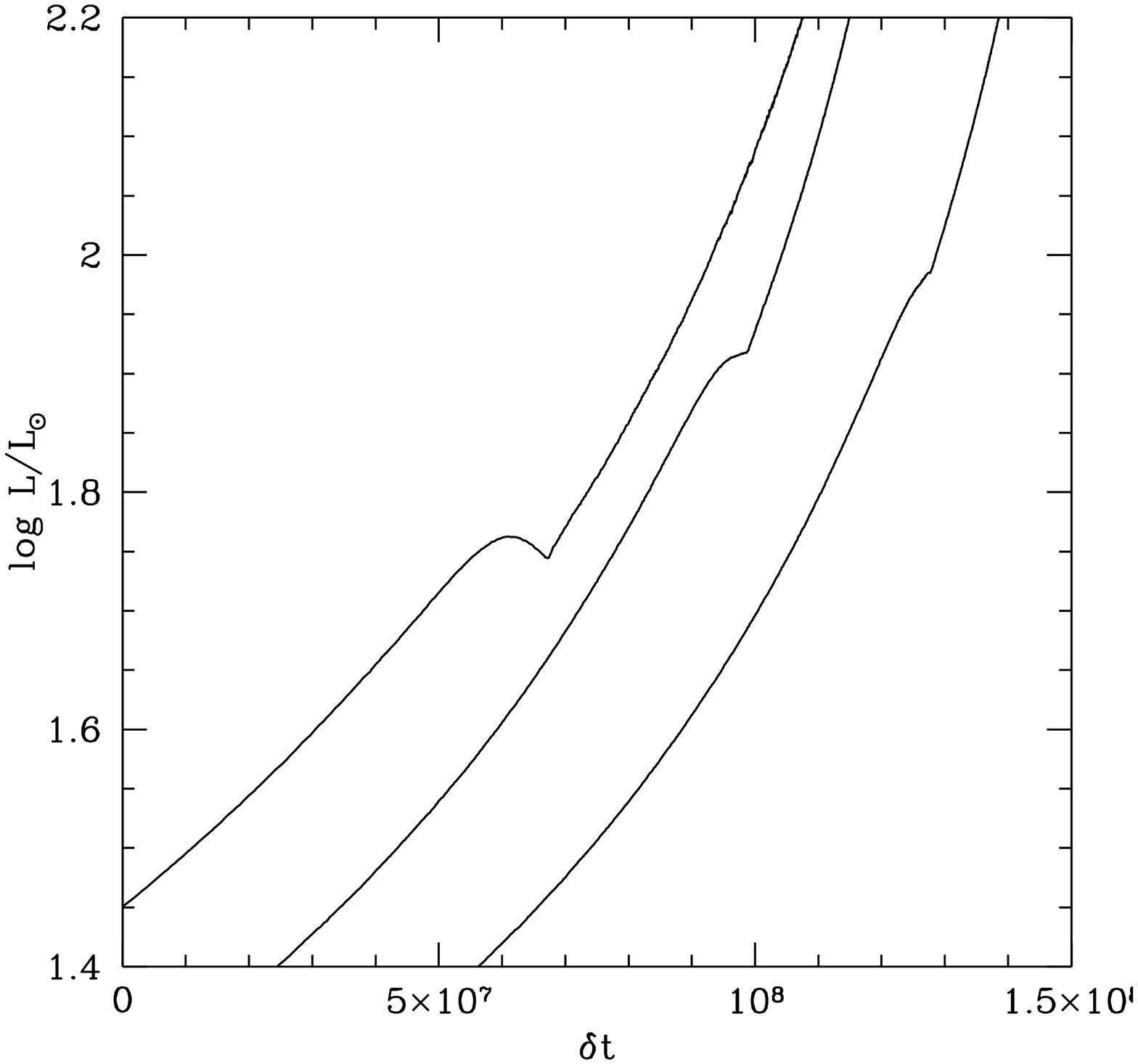}
 \caption{We show the luminosity versus time evolution of the three tracks (Y increasing from
 left to right). The times are arbitrarily normalized. Red giant bump of
 the Y=0.4 model is much less prominent than the others.}
\label{f3app}
\end{figure}

Fig.\ref{f1app} shows that the deepening of convection in the Y=0.24 model soon
produces a molecular weight discontinuity in the stellar envelope. The discontinuity
reaches a maximum at a mass fraction M$_r$/\Msun $\simeq 0.27$, due to the
maximum extension of convection at the so called ``first dredge up". Also a model
above the RG bump, when the burning shell has crossed the molecular weight
discontinuity, is shown. We see from the central and right figure
that the discontinuity is less prominent for the Y=0.35 and Y=0.40 models, and 
especially that it is not at all present during the first phases of evolution 
along the RGB. The location of the discontinuity is at about the same mass point
(M$_r$/\Msun $\simeq 0.27$) for the three cases. We show in Fig.\ref{f2app} the
HR diagram of the three tracks and the location at which the shell crosses the
discontinuity. For the Y=0.35 and Y=0.4 cases the effect on the 
luminosity evolution is barely seen in
the tracks, and does not produce any appreciable effect on the luminosity
function, as we can understand by looking at the time evolution of the luminosity,
shown in Fig.\ref{f3app}. As a byproduct, notice that this result shows that
we can not infer the existence of groups of stars very much enriched in initial
helium content from the effect on the RGB bump. 

Coming back to the paradigm of the extra--mixing in the RGB evolution, we 
hypothesize that the extreme helium rich models 
may not be able to avoid extra--mixing along the RGB, and also that extra mixing may 
not necessarily begin above the bump luminosity, as there is no or
a very small $\mu$ discontinuity which can prevent it.

\label{lastpage}

\end{document}